\documentclass{PoS}
\usepackage{epsfig}
\title{The LOFAR Transients Key Project}

\ShortTitle{The LOFAR Transients Key Project}

\author{Rob Fender,$^{ab}$ Ralph
         Wijers,$^b$ Ben Stappers,$^{bc}$ Robert Braun,$^c$ Michael Wise,$^{bc}$ Thijs Coenen,$^{b}$ Heino
         Falcke,$^{cd}$ Jean-Mathias Griessmeier,$^e$ Michiel van
         Haarlem,$^c$ Peter Jonker,$^{fgh}$ Casey Law,$^b$ Sera
         Markoff,$^b$ Joseph Masters,$^b$ James
         Miller-Jones,$^b$ Rachel Osten,$^i$ Bart Scheers,$^b$ Hanno
         Spreeuw,$^b$ John Swinbank,$^b$ Corina Vogt,$^c$ Rudy
         Wijnands$^b$ and Philippe Zarka,$^e$\\
        \llap{$^a$}School of Physics and Astronomy, University of
         Southampton, Highfield, Southampton, SO17 1BJ, UK\\
        \llap{$^b$}Astronomical Institute `Anton Pannekoek',
         University of Amsterdam, Kruislaan 403, 1098 SJ, Amsterdam,
         the Netherlands\\
        \llap{$^c$}Stichting ASTRON, Postbus 2, 7990 AA Dwingeloo, the
         Netherlands\\
        \llap{$^d$}Department of Astronomy, Radboud University,
         Postbus 9010, 6500 GL Nijmegen, the Netherlands\\
        \llap{$^e$}Observatoire de Paris-Meudon, 5 Place Jules
         Janssen, 92195 Meudon Cedex, France\\
        \llap{$^f$}SRON, Netherlands Institute for Space Research,
         Sorbonnelaan 2, 3584 CA, Utrecht, the Netherlands\\
        \llap{$^g$}Harvard-Smithsonian Center for Astrophysics, 60
         Garden Street, Cambridge, MA 02138, USA\\
        \llap{$^h$}Astronomical Institute, Utrect University, Postbus
         80000, 3508 TA, Utrecht, the Netherlands\\
        \llap{$^i$}Department of Astronomy, University of Maryland,
         College Park, MD, USA\\
        E-mail: \email{rpf@phys.soton.ac.uk}, \email{rwijers@science.uva.nl},\email{bws@science.uva.nl}, 
\email{braun@astron.nl},
                  \email{wise@science.uva.nl}, \email{tcoenen@science.uva.nl},
         \email{falcke@astron.nl}, \email{jean-mathias.griessmeier@obspm.fr},
         \email{haarlem@astron.nl}, \email{p.jonker@sron.nl},
         \email{claw@science.uva.nl}, \email{sera@science.uva.nl},
         \email{jmasters@science.uva.nl},
         \email{jmiller@science.uva.nl}, \email{rosten@astro.umd.edu},
         \email{bscheers@science.uva.nl},
         \email{hspreeuw@science.uva.nl}, \email{swinbank@science.uva.nl}
         \email{vogt@astron.nl}, \email{rudy@science.uva.nl},
         \email{philippe.zarka@obspm.fr}}

\abstract{LOFAR, the Low Frequency Array, is a new radio telescope
  under construction in the Netherlands, designed to operate between
  30 and 240\,MHz.  The Transients Key Project is one of the four Key
  Science Projects which comprise the core LOFAR science case.  The
  remit of the Transients Key Project is to study variable and
  transient radio sources detected by LOFAR, on timescales from
  milliseconds to years.  This will be achieved via both regular
  snapshot monitoring of historical and newly-discovered radio
  variables and, most radically, the development of a `Radio Sky
  Monitor' which will survey a large fraction of the northern sky on a daily basis.}

\FullConference{VI Microquasar Workshop: Microquasars and Beyond\\
		September 18-22 2006\\
		Societ\`a del Casino, Como, Italy}

\begin{document}

\section{LOFAR}

LOFAR (the {\bf LO}w {\bf F}requency {\bf AR}ray) is a new
radio telescope under construction in The Netherlands, which will
operate between 30 and 240\,MHz, providing unprecedented sensitivity
and resolution with which to explore the low-frequency radio sky.  It
will initially comprise 77 stations, 32 located in the 3\,km $\times
2$\,km core and the remaining 45 forming more extended baselines, with
a maximum baseline of about 100\,km.  The station locations have been
selected to provide the best possible {\it uv}-coverage both for
snapshot observations and for full aperture syntheses.  The stations
will return a total of 32\,MHz of bandwidth with a
basic integration time of 1\,s.  This 32 MHz of bandwidth may be
divided between multiple beams, e.g. 8 beams of 4 MHz each. Key
operational parameters are outlined in Table \ref{tab:lofar_parms}.
Each station will consist of 96 station units, each comprising one
low-band dipole (optimised for the range 30--80\,MHz) and a $4\times4$
array of high-band tiles (optimised for 120-240\,MHz).

\begin{table}[h]
\begin{tabular}{|cc|ccc|ccc|}
\hline
 && \multicolumn{3}{|c|}{Full array} & \multicolumn{3}{|c|}{24-beam Radio Sky Monitor}  \\
Frequency & A$_{\rm Eff}$ & $\Omega$ & $\theta$ & S & $\Omega$ & $\theta$ & S\\
(MHz) & (km$^2$) & (deg$^2$) & (arcsec) & (mJy\,bm$^{-1}$) & (deg$^2$) & (arcmin) & (mJy\,bm$^{-1}$) \\
\hline
30 & 0.19 & 120 & 25 & 118 & 2700 & 21 & 290 \\
75 & 0.03 & 20 & 10 & 80 & 420 & 8 & 200 \\
\hline
120 & 0.19 & 7 & 6 & 4 & 170 & 5 & 10 \\
200 & 0.07 & 3 & 3 & 4 & 60 & 3 & 9 \\
\hline
\end{tabular}
\caption{Key operational parameters of LOFAR, assuming a total
 bandwidth of 4\,MHz, an integration time of 1\,s and a maximum
 baseline of 100km. A$_{\rm Eff}$ is the effective area of the
 telescope, $\Omega$ is the field of view, and $\theta$ is the angular
 resolution.}
\label{tab:lofar_parms}
\end{table}

Four Key Science Projects have currently been identified for LOFAR,
each of which is associated with a university in The Netherlands:

\begin{table*}[h]
\begin{tabular}{lcr}
\hline\\
Key Project & Leading institution & Project Leader\\
\hline\\
The Epoch of Reionization & Groningen & de Bruyn \\
Extragalactic Surveys & Leiden & R\"ottgering\\
Transients  &Amsterdam & Fender/Wijers/Stappers\\
Cosmic Rays &Nijmegen & Kuijpers/Falcke\\
\hline\\
\end{tabular}
\end{table*}

In this brief article we will focus on the Transients Key Project.

\section{The Transients Key Science Project}

The Transients Key Science Project (TKP) aims to study all variable
sources detected by LOFAR.  The scientific remit of the project has been
subdivided into five basic categories: jet sources, pulsars (and
related neutron star phenomena), planets, flare stars and serendipity.

\begin{itemize}
\item{The class of jet sources includes all objects producing radio
emission from (often relativistic) outflows, including Active Galactic
Nuclei (AGN), Gamma Ray Bursts, and accreting white dwarfs, neutron
stars and `stellar mass' black holes (i.e. `microquasars').
Incoherent synchrotron radiation is the primary emission mechanism,
although there could be the possibility of detecting coherent prompt
emission from Gamma Ray Bursts.}
\item{A major survey of classical radio pulsars will be undertaken, as
well as the study of related objects such as Anomalous X-ray Pulsars
(AXPs) and Rotating Radio Transients (RRATs). LOFAR ill provide the
sensitivity to allow us to study the individual pulses from an
unprecedented number of pulsars including millisecond pulsars and the
bandwidth and frequency agility to study them over a wide range which
will provide vital new input for models of pulsar emission. }
\item{LOFAR will also study radio emission from planets within the Solar
System.  This includes imaging Jupiter's magnetosphere at high spatial
and time resolution, imaging Jupiter's radiation belts, and studying
planetary lightning from the other planets within the Solar System.
It is hoped that radio bursts from nearby so-called `hot Jupiter'
exoplanets might also be detected, and a survey will be carried out.}
\item{Flare stars and active binaries are likely to be present in almost
every LOFAR beam, giving off highly circularly-polarised radio bursts
from coherent emission processes.  Potential targets include M dwarf
flare stars, active binaries and low mass L and T dwarfs.}
\item{It is highly likely that probing such a large region of
hitherto-unexplored parameter space, we will detect new classes of
astrophysical objects. This possibility is especially enhanced by the
wide-field monitoring mode of the Radio Sky Monitor (see below).  We
aim to follow up such serendipitous discoveries both `internally' with
LOFAR and target-of-opportunity observations on other facilities, both
at higher radio frequencies and in other wavebands.}
\end{itemize}

\section{The radio sky monitor (RSM)}

\begin{figure*}[t]
\centerline{\epsfig{file=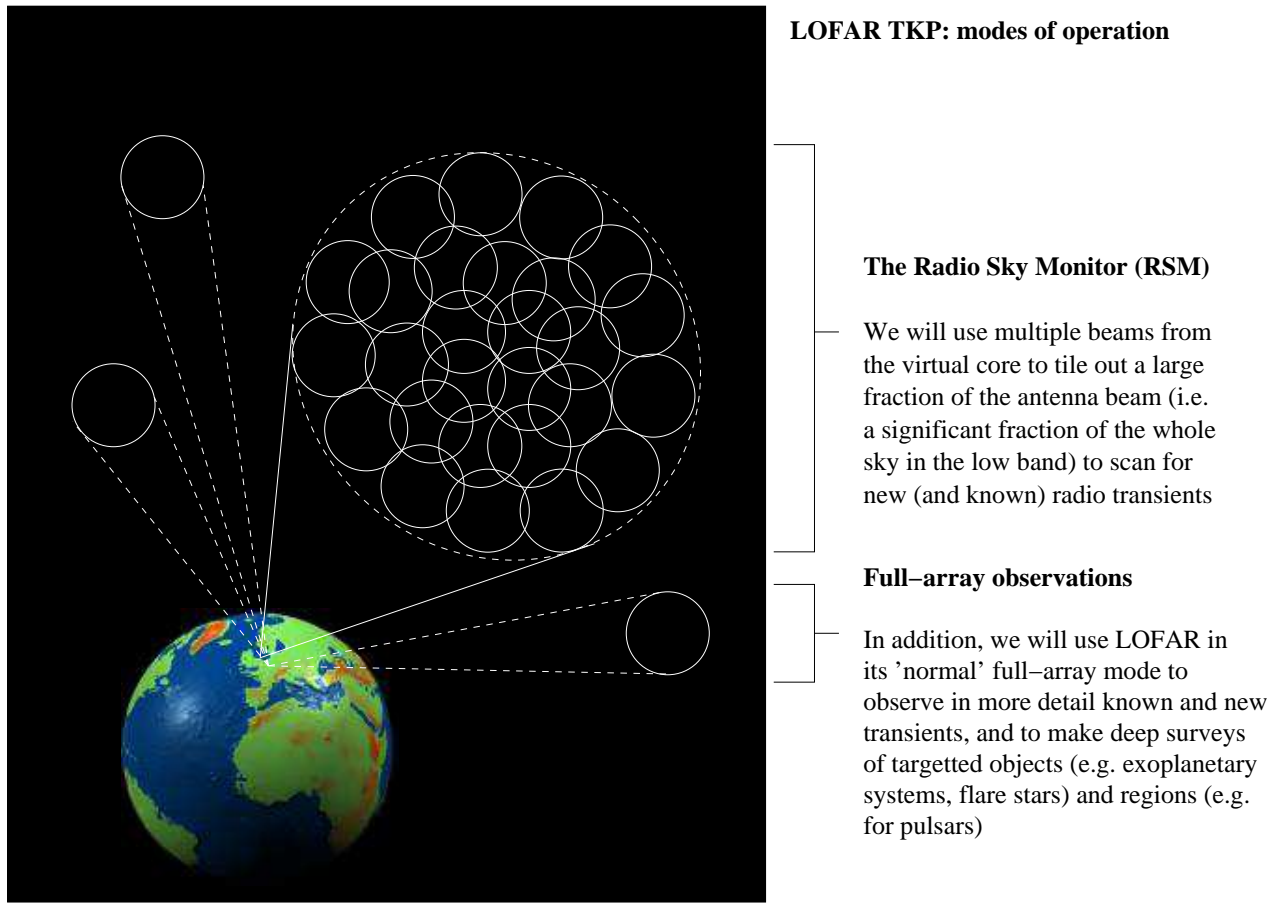, angle=0, width=17cm}}
\label{ASM-A}
\caption{Modes of operation of LOFAR for the Transients Key Project.
As well as utilising `normal' full-array modes, we plan to regularly
sweep the sky with a `Radio Sky Monitor' in which we use multiple
beams from the virtual core of the array to tile a large fraction
of the sky.}
\end{figure*}

LOFAR is a software telescope.  The individual dipoles are sensitive
to the entire hemisphere of sky above them, but by dividing up the
total bandwidth and electronically inserting delays in the signal
path, multiple beams may be created to give simultaneous directional
sensitivity in several different directions.

The wide field of view of the compact LOFAR core is ideal for
observing large areas of sky.  The 32 core stations will ultimately be
able to return up to 24 individual beams (each of reduced total
bandwidth), which will tile out a patch of sky in a radio sky monitor
mode. The total area tiled out will be a sizeable fraction of the
individual antenna response, i.e. a large fraction of the sky at the
lowest frequencies. In parallel to this, depending on useage of
computer resources, other beams using all LOFAR stations (core and
remote stations) may be able to simultaneously carry out separate
observing programmes.

Thus, for the first time, it will be possible to have a Radio Sky
Monitor (RSM) which will be able to observe more or less the entire
sky visible from the Netherlands on a daily basis.  The results of
this monitoring programme will be made available to the community in
the form of alerts when new transients are detected and real-time
lightcurves for most sources under observation.  We will create and
maintain a database of all known transient radio sources, constantly
updated as new observations are made. Bright transients may result in
internal triggers allowing full-array observations and arcsec-accuracy
localisation within tens of seconds. We envisage this mode as being a
key resource for the discovery and localisation of new transients for
the high-energy astrophysics community. Furthermore, this mode may act
as a prototype for a low-frequency core of the Square Kilometer Array.

\section{The transient buffer boards: looking back in time}

Sufficient memory will be available to record one second of raw data
for each LOFAR antenna at full bandwidth.  By trading bandwidth for
time, the length of time stored in the buffer could potentially be
increased by a large factor.  On detection of a LOFAR transient, or on
reception of a trigger from an external telescope, the data in the
buffers will be frozen, and fed back to the processing cluster, to
allow us to go `back in time' and take a more detailed look at the
onset of the transient event, or to probe the event on shorter
timescales. 

\section{Strategy}

The sky will be searched for transients by making differenced images
on a logarithmically-spaced range of timescales, ranging (initially)
from 1\,s up to $10^4$\,s, thus probing a wide variety of transient
events (targetted observations of pulsars, flare stars and planets
will necessarily be made with much higher time resolution).  In order
to provide real-time transient alerts, and if necessary reconfigure
the telescope to better localise a new transient, all data processing
must be fully automated.  The automated pipeline must keep pace with
the incoming data, taking no more than 1\,s to process the
shortest-timescale maps.  Each timestep, a difference image will be
made comparing the current image with the previous one.  A source
extraction routine will be run on the differenced image to detect new
transients.  Any sources detected will be subjected to RFI (radio
frequency interference) and noise checking, and measurements of the
source parameters (flux density, spectrum, polarisation and position)
will be made. When searching for very rapid events, real-time
de-dispersion must also be applied.  After comparison with the
existing database, new sources will be classified using a
probability-based scheme based on the measured parameters, and the
parameters of known sources will be used to update the monitoring
database.  New transients may be followed up using pointed LOFAR
observations and target of opportunity proposals at external
observatories.  In order to probe shorter timescale events, a
tied-array mode will be available to provide 5\,$\mu$s time resolution
to enable the study of pulsars, planetary bursts, and other sources of
coherent emission.

\section{Timeline}
The first LOFAR station (Core Station 1; CS\,1) is already in place
and is undergoing commissioning.  Once the scientific testing phase
begins, we will be in a position to detect a bright transient event
with CS\,1.  The central LOFAR core will be constructed over the
course of 2007, increasing the available sensitivity and bringing the
RSM online.  It is hoped that long-baseline international stations (in
Germany, The UK, France and beyond) will begin to be added at this
stage.  The full array is planned to be completed by the end of 2008,
with science operations beginning in early 2009.  LOFAR will
ultimately be fully international and open time will be available via
competitive merit-based proposals.

\end{document}